\documentclass[a4paper,aps,prd,twocolumn,preprintnumbers,showpacs,superscriptaddress,nofootinbib]{revtex4}
\usepackage{graphics,graphicx,epsfig}
\usepackage{amsfonts,amsmath,amssymb}

\input{tcilatex}

\begin{document}

\title{Quasilocal energy for three-dimensional massive gravity solutions
with chiral deformations of AdS$_3$ boundary conditions}
\author{Alan Garbarz}
\email{alan-at-df.uba.ar}
\affiliation{Departamento de F\'{\i}sica, Universidad de Buenos Aires FCEN-UBA,
IFIBA-CONICET, Ciudad Universitaria, Pabell\'on I, 1428, Buenos Aires,
Argentina.}
\affiliation{Instituto de F\'{\i}sica de La Plata, Universidad Nacional de La Plata
IFLP-UNLP, C.C. 67, 1900, La Plata, Argentina.}
\author{Gaston Giribet}
\email{gaston-at-df.uba.ar}
\affiliation{Departamento de F\'{\i}sica, Universidad de Buenos Aires FCEN-UBA,
IFIBA-CONICET, Ciudad Universitaria, Pabell\'on I, 1428, Buenos Aires,
Argentina.}
\author{Andr\'es Goya}
\email{af.goya-at-df.uba.ar}
\affiliation{Departamento de F\'{\i}sica, Universidad de Buenos Aires FCEN-UBA,
IFIBA-CONICET, Ciudad Universitaria, Pabell\'on I, 1428, Buenos Aires,
Argentina.}
\author{Mauricio Leston}
\email{mauricio-at-iafe.uba.ar}
\affiliation{Instituto de Astronom\'{\i}a y F\'{\i}sica del Espacio IAFE-CONICET, Ciudad
Universitaria, C.C. 67 Suc. 28, 1428, Buenos Aires, Argentina.}

\begin{abstract}
We consider critical gravity in three dimensions; that is, the New Massive
Gravity theory formulated about Anti-de Sitter (AdS) space with the specific
value of the graviton mass for which it results dual to a two-dimensional
conformal field theory with vanishing central charge. As it happens with
Kerr black holes in four-dimensional critical gravity, in three-dimensional
critical gravity the Ba\~{n}ados-Teitelboim-Zanelli black holes have
vanishing mass and vanishing angular momentum. However, provided suitable
asymptotic conditions are chosen, the theory may also admit solutions
carrying non-vanishing charges. Here, we give simple examples of exact
solutions that exhibit falling-off conditions that are even weaker than
those of the so-called Log-gravity. For such solutions, we define the
quasilocal stress-tensor and use it to compute conserved charges. Despite
the drastic deformation of AdS$_3$ asymptotic, these solutions have non-zero
mass and angular momentum, which we compute.
\end{abstract}

\maketitle

\section{Introduction}

In three dimensions a fully covariant parity-even theory of gravity that
reduces to massive spin-two Fierz-Pauli theory at linearized level does
exist. This is the so-called New Massive Gravity theory (NMG), which was
proposed in Ref. \cite{NMG, NMG2}. When formulated about asymptotically
Anti-de Sitter (AdS) spacetime, NMG exhibits features that are reminiscent
of those of cosmological Topologically Massive Gravity (TMG) \cite{TMG, TMG2}%
. In particular, a special point of the parameter space that resembles the
chiral point of TMG exists in NMG as well. At this point, the central charge
of the dual two-dimensional conformal field theory (CFT) vanishes \cite%
{GrumillerHohm} and, as it happens in TMG \cite{GrumillerJohansson}, the
asymptotic boundary conditions may be relaxed with respect to the standard
Brown-Henneaux boundary conditions \cite{BH}. The aim of the present paper
is to study how the asymptotic conditions may be relaxed.

Asymptotic boundary conditions are of central importance in
three-dimensional gravity, particularly in the case of negative cosmological
constant. For the theory on AdS$_{3}$, boundary conditions are crucial to
realize the action of the conformal group at the boundary of the spacetime 
\cite{BH}, what is ultimately interpreted in a natural way within the
context of AdS/CFT correspondence \cite{Malda}. A good example to illustrate
in what sense boundary conditions are essential to define the theory is
Chiral Gravity: A few years ago, when Chiral Gravity was proposed in Ref. 
\cite{ChiralGravity}, the discussion about the appropriate boundary
conditions to be imposed was the key point to determine the consistency of
the model \cite{CDWW, CDWW2, GKP}. It was shown in Ref. \cite{LogGravity}
how much the properties of TMG at the chiral point depend on the asymptotic
conditions considered. In fact, the theory presents quite distinct features
depending on whether one considers the orthodox boundary conditions
originally proposed in \cite{BH} or, on the contrary, one opts for the
weakened version proposed in \cite{GrumillerJohansson, GrumillerJohansson2}.
In the former case, the bulk theory results to be dual to an holomorphic
CFT, whose symmetry is generated by a single copy of Virasoro algebra. In
contrast, in the latter case, TMG\ at the chiral point results to be dual to
a non-unitary CFT whose symmetry is generated by the product of a Virasoro
algebra and a Witt algebra \cite{Henneaux1}. Other choices of boundary
conditions leading to different symmetry algebras at the boundary were also
considered in the literature, and the discussions about this point are
interesting and extend to different three-dimensional models, including TMG,
NMG and Einstein gravity coupled to matter; see for instance \cite{Zanelli,
OTT, Henneaux2, Sun1, Sun2, Sun3, NewBC}.

In this paper, we will consider solutions obeying a set of boundary
conditions different from those of \cite{BH} and \cite{GrumillerJohansson}.
The asymptotic conditions studied here turn out to be a deformation of
Brown-Henneaux boundary conditions, but of a different type since they not
only relax the next-to-leading components of the metric but also some
leading pieces of it. Remarkably, although the conditions we will consider
change the asymptotic behavior drastically, quasilocal stress tensor can
still be consistently defined at the boundary of the spacetime \cite%
{BrownYork} and be used to compute conserved charges associated to exact
solutions obeying the new boundary conditions. The quasilocal energy for
deformed AdS$_{3}$ boundary conditions was computed, for instance, in Refs. 
\cite{HohmTonni, GiribetLeston, Goya1, Troncoso, Goya2}; however, as said,
the asymptotic conditions considered here exhibit a substantially weaker
damping-off.

The paper is organized as follows:\ In section 2, we will review
three-dimensional massive gravity. We will be involved with the theory
proposed in \cite{NMG} formulated on AdS$_{3}$ space at the point of the
parameter space at which the central charge of the dual CFT vanishes. This
amounts to tune the value of the graviton mass in terms of the effective
cosmological constant. In section 3, with the aim of presenting the set of
boundary conditions we are interested in, we will first summarize different
choices of asymptotic behaviors considered in the literature and compare
their falling-off conditions in the near boundary limit. In section 4, we
will present a simple example of an exact solution that satisfies the new
boundary conditions but can not be accommodated neither in the
Brown-Henneaux nor in Log-gravity boundary conditions. The particular
solution we will discuss is a chiral deformation of the extremal Ba\~{n}%
ados-Teitelboim-Zanelli (BTZ) black hole \cite{BTZ, BTZ2}. It generalizes
explicit solutions to TMG and NMG previously found in Refs. \cite{Gibbons,
GAY, Clement}. For these solutions, we will show in section 5 that
quasilocal stress-tensor can be consistently defined in such a way that both
the mass and angular momentum of the deformed BTZ hole can be computed.
Remarkably, despite the drastic deformation of AdS$_{3}$ asymptotic behavior
we consider, the solutions have finite mass and angular momentum, which we
compute.

\section{Massive Gravity}

\subsection{The Fierz-Pauli action}

Let us start by reviewing massive gravity in $D$ dimensions. This is given
by the Fierz-Pauli action%
\begin{eqnarray}
S &=&\int d^{D}x\left( -\frac{1}{2}\partial _{\lambda }h_{\mu \nu }\partial
^{\lambda }h^{\mu \nu }+\partial _{\mu }h_{\nu \lambda }\partial ^{\nu
}h^{\mu \lambda }-\right.  \notag \\
&&\left. \partial _{\mu }h^{\mu \nu }\partial _{\nu }h+\frac{1}{2}\partial
_{\lambda }h\partial ^{\lambda }h-\frac{m^{2}}{2}(h_{\mu \nu }h^{\mu \nu
}-h^{2})\right)  \label{FP}
\end{eqnarray}%
for the symmetric field $h_{\mu \nu }$ that represents the gravitational
perturbation about Minkowski space, $g_{\mu \nu }=\eta _{\mu \nu }+\kappa
h_{\mu \nu }$, with $\kappa ^{2}=16\pi l_{P}^{D-2}$. As usual, we denote $%
h\equiv \eta ^{\mu \nu }h_{\mu \nu }$. Here we will set the Planck length $%
l_{P}$ to one.

The kinetic term in (\ref{FP}) coincides with the Einstein-Hilbert action at
second order in $h_{\mu \nu }$. The relative factor $-1$ in the massive term
is chosen for the component $h^{00}$ of the metric to appear as a Lagrange
multiplier; then the theory has only five local degrees of freedom in $D=4$
dimensions and two degrees of freedom in $D=3$ dimensions.

The theory may also be formulated about other backgrounds. Specially about
(A)dS spaces, for which one has to introduce a cosmological term to the
action above. Nevertheless, let us first discuss the theory about Minkowski
space and introduce the cosmological constant later, after discussing the
covariant extension of the theory.

The equations of motion derived from (\ref{FP}) are%
\begin{eqnarray}
&& \square h_{\mu \nu }-\partial _{\lambda }\partial _{\mu }h_{\nu
}^{\lambda }-\partial _{\lambda }\partial _{\nu }h_{\mu }^{\lambda }+\eta
_{\mu \nu }\partial _{\lambda }\partial _{\sigma }h^{\lambda \sigma }+ \\
&& \partial _{\mu }\partial _{\nu }h-\eta _{\mu \nu }\square h-m^{2}(h_{\mu
\nu }-\eta _{\mu \nu }h)=0.  \notag  \label{Dosa}
\end{eqnarray}

Here, the kinetic term coincides with Einstein tensor for metric $g_{\mu \nu
}$ at first order in $h_{\mu \nu }$. For this reason, we prefer to denote it 
$G_{\mu \nu }^{(1)}[h]$; namely 
\begin{eqnarray}
G_{\mu \nu }^{(1)}[h] &\equiv &\square h_{\mu \nu }-\eta _{\mu \nu }\square
h-\partial _{\lambda }\partial _{\mu }h_{\nu }^{\lambda }-\partial _{\lambda
}\partial _{\nu }h_{\mu }^{\lambda }+  \notag \\
&&\eta _{\mu \nu }\partial _{\lambda }\partial _{\sigma }h^{\lambda \sigma
}+\partial _{\mu }\partial _{\nu }h,  \label{G1}
\end{eqnarray}%
where the superindex $(1)$ refers to the linear nature.

Now, let us go back to equation (2). Acting on it with the differential
operator $\partial ^{\mu }$, if $m\neq 0$, one gets $\partial ^{\mu }h_{\mu
\nu }=\partial _{\nu }h.$ Plugging this back into (\ref{Dosa}) one finds%
\begin{equation}
\square h_{\mu \nu }-\partial _{\mu }\partial _{\nu }h-m^{2}(h_{\mu \nu
}-\eta _{\mu \nu }h)=0.  \label{3}
\end{equation}%
Taking the trace of (\ref{3}) one finds that the trace vanishes, $\eta ^{\mu
\nu }h_{\mu \nu }\equiv h=0$. This implies $\partial ^{\mu }h_{\mu \nu }=0$
and consequently, from (\ref{Dosa}), one gets the Klein-Gordon equation%
\begin{equation}
\square h_{\mu \nu }=m^{2}h_{\mu \nu }.  \label{5}
\end{equation}%
This, supplemented by $\partial ^{\mu }h_{\mu \nu }=0$ and $h=0,$ completes
the set of equations of a massive spin-two field about Minkowski spacetime.

\subsection{Massive gravity in three dimensions}

Now, let us discuss a curious feature that occurs in three dimensions.
Consider the particular case $D=3$ of theory (\ref{FP}); namely, consider
the system of equations%
\begin{equation}
(\square -m^{2})h_{\mu \nu }=0,\quad \partial ^{\mu }h_{\mu \nu }=0,\quad
\eta ^{\mu \nu }h_{\mu \nu }=0.  \label{8}
\end{equation}%
Next, replace the metric $h_{\mu \nu }$ by the linearized Einstein tensor
constructed out of it;\ namely, change%
\begin{equation}
h_{\mu \nu }\rightarrow G_{\mu \nu }^{(1)}[h]  \label{antes}
\end{equation}%
in (\ref{8}). As we will see, this simple trick will lead to an interesting
model that is only possible in three dimensions. After performing (\ref%
{antes}), the equations of motion turn out to be%
\begin{equation}
(\square -m^{2})G_{\mu \nu }^{(1)}[h],\quad R^{(1)}[h]=0,  \label{system}
\end{equation}%
where $R^{(1)}[h]$ represents the Ricci scalar associated to metric $g_{\mu
\nu }$ at first order in the perturbation $h_{\mu \nu }$. Then, taking into
account (\ref{G1}), the first of these equations can be written as 
\begin{equation}
G_{\mu \nu }^{(1)}[G^{(1)}[h]]=G_{\mu \nu }^{(1)}[m^{2}h],  \label{torba}
\end{equation}%
since, provided $\partial ^{\mu }h_{\mu \nu }=0$ and $h=0$, it holds $G_{\mu
\nu }^{(1)}[h]=\square h_{\mu \nu }$. In this way, one obtains a equations
that are invariant under linearized diffeomorphisms.

Now, an important observation:\ The key point to understand the peculiarity
of the three-dimensional case is to be reminded that in three dimensions
there exists a direct connection (local identification) between Einstein
tensor associated to a given metric and the metric itself. In fact, all
solutions to three-dimensional Einstein equations are locally equivalent, so
that, in what local information regards to, $g_{\mu \nu }$ and $G_{\mu \nu }$
carry exactly the same information. In particular, this implies that (\ref%
{torba}) expresses the local identify between $G_{\mu \nu }^{(1)}$ and $%
m^{2}h_{\mu \nu }$, which actually justifies having done (\ref{antes}).

Summarizing, something remarkable has been achieved:\ A theory has been
obtained that is invariant under linear diffeomorphisms and, at the same
time, is equivalent to massive Fierz-Pauli theory \cite{NMG}. This is due to
the magic of $D=3$ dimensions, c.f. \cite{NMG4D}. Below we will see how to
extend this theory in a fully covariant way.

An important observation is that system (\ref{system}) can be derived from
the following action%
\begin{eqnarray}
S_{\text{NMG}}^{(1)} &=&\frac{1}{16\pi }\int d^{3}x\left( \frac{1}{2}h^{\mu
\nu }G_{\mu \nu }^{(1)}[h]-\frac{1}{4m^{2}}G_{\mu \nu }^{(1)}[h]\right. 
\notag \\
&&\left. \left( R^{(1)\mu \nu }[h]-\frac{1}{4}\eta ^{\mu \nu
}R^{(1)}[h]\right) \right) ,  \label{NMGlin}
\end{eqnarray}
varying this action with respect to the field $h_{\mu \nu }$.

Then, it is easy to propose a generally covariant extension of (\ref{NMGlin}%
). Noticing that contracting the Einstein tensor $G_{\mu \nu }\equiv G_{\mu
\nu }^{(1)}+G_{\mu \nu }^{(2)}+...\equiv R_{\mu \nu }-(1/2)Rg_{\mu \nu }$
and the Schouten tensor $S_{\mu \nu }\equiv R_{\mu \nu }-(1/4)Rg_{\mu \nu }$
in three dimensions yields $G_{\mu \nu }S^{\mu \nu }=R_{\mu \nu }R^{\mu \nu
}-(3/8)R^{2}$, one can write the NMG\ action \cite{NMG}

\begin{equation}
S_{\text{NMG}}=\frac{1}{16\pi }\int_{\Sigma }d^{3}x\sqrt{-g}\left( R-\frac{1%
}{m^{2}}(R_{\mu \nu }R^{\mu \nu }-\frac{3}{8}R^{2})\right) ,  \label{NMG1}
\end{equation}%
which represents a fully covariant parity-even theory of massive gravity.

The equations of motion derived from action (\ref{NMG1}) read 
\begin{equation}
G_{\mu \nu }-\frac{1}{2m^{2}}K_{\mu \nu }=0,  \label{eom}
\end{equation}%
which, apart from the Einstein tensor $G_{\mu \nu }$, involve the tensor%
\begin{eqnarray}
K_{\mu \nu } &=&2\square {R}_{\mu \nu }-\frac{1}{2}\nabla _{\mu }\nabla
_{\nu }{R}-\frac{1}{2}\square {R}g_{\mu \nu }-\frac{3}{2}RR_{\mu \nu }- 
\notag \\
&&-R_{\alpha \beta }R^{\alpha \beta }g_{\mu \nu }+\frac{3}{8}R^{2}g_{\mu \nu
}+4R_{\mu \alpha \nu \beta }R^{\alpha \beta }.  \label{eom2}
\end{eqnarray}

These are fourth-order differential equations for the metric $g_{\mu \nu }$.
The precise combination of the square-curvature terms in (\ref{NMG1})
satisfies the notable property of being the trace of the equations it leads
to. That is, $g^{\mu \nu }K_{\mu \nu }=R_{\mu \nu }R^{\mu \nu }-\frac{3}{8}%
R^{2}$. This combination also happens to be the one that makes the
dependence $\square {R}$ to disappear from the trace of the equations of
motion. This results in the decoupling of a ghostly mode of higher-curvature
gravity. In turn, the theory contains (no more than) two local degrees of
freedom, which can be associated to a massive spin-2 field in $D=3$.

\subsection{The theory on Anti-de Sitter}

NMG action (\ref{NMG1}) may be supplemented with a cosmological constant
term 
\begin{equation}
S_{\Lambda }=-\frac{\Lambda }{8\pi }\int d^{3}x\sqrt{-g}.  \label{cosmo}
\end{equation}

The theory admits AdS$_{3}$ vacua provided the effective cosmological
constant $\Lambda _{\text{eff}}\equiv - l^{-2}$ is negative: Asking for
constant curvature solutions $R_{\mu \nu }=-(2/l^{2})g_{\mu \nu },$ one
finds $- \Lambda _{\text{eff}}\equiv l_{\pm }^{-2}=-(\Lambda /2)(1\pm \sqrt{%
1-\Lambda /m^{2}})$. Generically, this gives two values for the effective
cosmological constant, which we denote $l_{-}^{2}$ and $l_{+}^{2} $.
Provided one of these values is negative, AdS$_{3}$ space appears as a
solution, together with all the other geometries that are locally equivalent
to it, like the notable case of the BTZ black hole \cite{BTZ}.

The metric of AdS$_{3}$ spacetime can be written as%
\begin{equation}
ds_{\text{AdS}}^{2}=-\left( \frac{r^{2}}{l^{2}}+1\right) dt^{2}+\left( \frac{%
r^{2}}{l^{2}}+1\right) ^{-1}dr^{2}+r^{2}d\phi ^{2},  \label{AdS}
\end{equation}%
with $r\in \mathbb{R}_{\geq 0}$, $t\in \mathbb{R}$, $\phi \in \lbrack 0,2\pi
)$. The boundary of the space is located at $r=\infty $.

According to AdS/CFT conjecture, if NMG on AdS$_{3}$ resulted to be a
consistent model, then it should be dual to a two-dimensional CFT formulated
at $r=\infty $. Say this is actually the case. Then, it is expected that the
central charge of the dual CFT$_{2}$ will coincide with the central
extension of the algebra that generates the asymptotic isometry group. Its
value can easily be computed to be%
\begin{equation}
c=\frac{3l}{2}\left( 1-\frac{1}{2l^{2}m^{2}}\right)  \label{c}
\end{equation}%
which, indeed, agrees with the trace anomaly of the dual CFT$_{2}$ \cite%
{Goya1}.

In addition to $S_{NMG}+S_{\Lambda }$, the theory may be augmented by adding
to the action a gravitational Chern-Simons term \cite{TMG}%
\begin{equation}
S_{\text{CS}}=\frac{1}{32\pi \mu }\int_{\Sigma }d^{3}x\varepsilon ^{\mu \nu
\rho }\Gamma _{\mu \alpha }^{\beta }\left( \partial _{\nu }\Gamma _{\rho
\beta }^{\alpha }-\frac{2}{3}\Gamma _{\nu \delta }^{\alpha }\Gamma _{\rho
\beta }^{\delta })\right) .  \label{CS}
\end{equation}

In such case, equations of motion (\ref{eom}) acquires an additional term
proportional to the Cotton tensor $C_{\mu \nu }=\frac{1}{2}\varepsilon _{\mu
}^{\ \text{\ }\alpha \beta }\nabla _{\alpha }R_{\beta \nu }+\frac{1}{2}%
\varepsilon _{\nu }^{\ \text{\ }\alpha \beta }\nabla _{\alpha }R_{\mu \beta
} $ with a coupling constant $1/\mu $. Equation (\ref{c}) is also modified
by the inclusion of (\ref{CS}): For finite $\mu $ the dual CFT$_{2}$
exhibits diffeomorphism anomaly and thus the right-moving $c_{+}$ and
left-moving $c_{-}$ central charges are different;\ more precisely, $c_{\pm
}=c\pm 3/(2\mu )$, with $c$ given by (\ref{c}). Chiral Gravity theory of 
\cite{ChiralGravity} corresponds to $\mu l=1$ and $1/m^{2}=0$, and can be
naturally generalized to finite $m^{2}$ by demanding $c_{-}=0$. Here, we
will be mainly interested in the case $1/\mu =0$ at the point of the
parameter space where $c=0$, where the graviton mass is%
\begin{equation}
m^{2}=\frac{1}{2l^{2}}.  \label{aster}
\end{equation}%
At this point, NMG exhibits peculiar features. For instance, when (\ref%
{aster}) holds all BTZ black holes have vanishing conserved charges. It is
worth mentioning that, despite of this fact, the theory at $c=0$ also
presents solutions with non-vanishing conserved charges \cite{GiribetLeston}%
. We will discuss solutions of this sort.

\section{Boundary conditions}

\subsection{Asymptotic boundary conditions}

As we mentioned in the introduction, an important ingredient to define the
theory are the boundary conditions. Standard Brown-Henneaux boundary
conditions are specified by considering deformations of AdS$_{3}$ metric (%
\ref{AdS}) of the form 
\begin{equation}
g_{\mu \nu }=g_{\mu \nu }^{(\text{AdS})}+h_{\mu \nu }
\end{equation}%
and demanding the components of $h_{\mu \nu }$ to damp off in the following
manner \cite{BH}%
\begin{eqnarray}
h_{tt} &\simeq &\mathcal{O}(1),\quad h_{rr}\simeq \mathcal{O}(r^{-4}),\quad
h_{\phi t}\simeq \mathcal{O}(1),  \label{bh} \\
h_{\phi r} &\simeq &\mathcal{O}(r^{-3}),\quad h_{\phi \phi }\simeq \mathcal{O%
}(1),\quad h_{rt}\simeq \mathcal{O}(r^{-3})  \label{bh2}
\end{eqnarray}%
where $\mathcal{O}(r^{-n})$ stands for arbitrary functions of coordinates $%
\phi $ and $t$ that fall off equally or faster than a power $r^{-n}$ at
large $r$. In particular, this implies%
\begin{equation}
g_{tt}=-\frac{r^{2}}{l^{2}}+\mathcal{O}(1),\quad g_{\phi \phi }=r^{2}+%
\mathcal{O}(1).  \label{25}
\end{equation}%
This set of boundary conditions incorporates, in particular, the BTZ black
hole solutions.

Log-gravity boundary conditions proposed in \cite{GrumillerJohansson,
GrumillerJohansson2} permit relaxation of (\ref{25}) including terms like $%
h_{ij}\simeq \mathcal{O}(\log (r))$ with $i,j=t,\phi $. Also, other sets of
boundary conditions can be defined \cite{Sun3, Waves, OTT, Henneaux2}. In
order to clearly distinguish between different boundary conditions, let us
summarize some proposals below. To do this, consider the following form of
the metric%
\begin{equation}
ds^{2}=d\rho ^{2}+e^{2\rho }\gamma _{ab}dz^{a}dz^{b},  \label{meee}
\end{equation}%
which resembles the Fefferman-Graham expansion of General Relativity \cite%
{FG}, with $z^{\pm }$ being two null directions ($a,b=\pm $), with the
asymptotic expansion%
\begin{equation}
\gamma _{ab}(\rho )=\gamma _{ab}^{(0)}+e^{-2\rho }\gamma
_{ab}^{(2)}+e^{-4\rho }\gamma _{ab}^{(4)}+...  \label{carita}
\end{equation}%
where $\gamma _{ab}^{(n)}$ are functions of $z^{+}$ and/or $z^{-}$ that do
not depend on $\rho $; see for instance Ref. \cite{Cunliff}.

In terms of (\ref{meee})-(\ref{carita}), Brown-Henneaux boundary conditions (%
\ref{bh})-(\ref{bh2}) read 
\begin{equation}
\gamma _{--}^{(0)}=\gamma _{++}^{(0)}=0,\quad \gamma _{-+}^{(0)}=\gamma
_{+-}^{(0)}=-\frac{1}{2}.  \label{c1}
\end{equation}%
To compare with (\ref{bh})-(\ref{bh2}) consider the change of coordinates $%
\tau \equiv t/l^{2}$, $\varphi \equiv \phi /l$, with $z^{\pm }\equiv \tau
\pm \varphi $ and $\rho \equiv \log (r)$.

The so-called Log-gravity relaxed boundary conditions \cite%
{GrumillerJohansson} correspond to considering NMG at the point $%
m^{2}l^{2}=1/2$ and supplementing expansion (\ref{carita}) with an
additional term $\rho e^{-2\rho }\gamma _{++}^{(\text{Log})}$ keeping (\ref%
{c1}). Provided NMG is parity-even, a ($--$) version of these conditions
also exists \cite{Sun2}. Clearly, the asymptotic behavior with $\gamma
_{++}^{(\text{Log})}\neq 0$ is weaker than Brown-Henneaux conditions due to
the existence of a term in (\ref{meee}) that grows linearly as $\sim 
\mathcal{O}(\rho )$. In terms of radial coordinate $r$ this corresponds to a
logarithmic term $\sim \mathcal{O}(\log (r))$.

In Ref. \cite{Sun3}, the possibility of having contributions to (\ref{meee})
that grow quadratically as $\sim \mathcal{O}(\rho ^{2})\sim \mathcal{O}(\log
^{2}(r))$ was considered. It was shown that such a behavior is possible if
NMG is coupled to TMG\ at the fine tuned point $ml^{2}=-2\mu l=-3/2.$ In
such case, it is possible to supplement (\ref{carita}) with a term $\rho
^{2}e^{-2\rho }\gamma _{++}^{(\text{Log}^{2})}$ with (\ref{c1}). Explicit
solutions obeying these conditions were found in Ref. \cite{Waves}.

A totally different set of boundary conditions for NMG\ coupled to TMG\ was
proposed in Ref. \cite{OTT}, where it was shown that the theory\ at the
point $m^{2}l^{2}=-1/2$ with arbitrary $\mu $ admits to add to (\ref{carita}%
) a term like $e^{-\rho }\gamma _{+-}^{(1)}.$ This is a term that grows
quite rapidly at large $\rho $, as it gives a next-to-leading contribution $%
\sim \mathcal{O}(e^{\rho })\sim \mathcal{O}(r)$ to (\ref{meee}). The
computation of quasilocal stress-tensor of solutions satisfying this
asymptotic and/or the Log-gravity boundary conditions was done, for
instance, in Refs. \cite{HohmTonni, GiribetLeston}.

It is interesting to compare the behaviors listed above with the new type of
boundary conditions proposed by Comp\`{e}re, Song, and Strominger in \cite%
{NewBC} in the context of three-dimensional Einstein gravity coupled to
matter. These correspond to considering expansion (\ref{carita}) and
relaxing (\ref{c1}) by allowing for $\gamma _{++}^{(0)}=\partial
_{+}f(z^{+})\neq 0$ with $\gamma _{-+}^{(0)}=\gamma _{+-}^{(0)}=-1/2$, and $%
\gamma _{--}^{(2)}$ fixed. These boundary conditions were shown to yield an
asymptotic isometry algebra generated by the product of a single Virasoro
algebra and an affine Kac-Moody $\widehat{u}(1)$ factor. Reduced conformal
symmetry for TMG on AdS$_{3}$ with boundary conditions of mixed chirality
was also studied in Ref. \cite{Henneaux2}.

Here, we will consider a different set of boundary conditions. We will
consider deformations of Brown-Henneaux asymptotic (\ref{c1}) that
correspond to supplementing expansion (\ref{carita}) with terms 
\begin{equation}
\rho e^{-2\rho }\gamma _{++}^{(\text{Log})}+\rho \gamma _{++}^{(\text{New})}.
\label{c6}
\end{equation}%
These asymptotic conditions reduce to Log-gravity conditions only in the
case $\gamma _{++}^{(\text{New})}=0$, while in the case such a term is
turned on the metric (\ref{meee}) acquires a dependence like $\sim \mathcal{O%
}(\rho e^{2\rho })\sim \mathcal{O}(r^{2}\log (r))$. Notice that, in contrast
to Brown-Henneaux and Log-gravity boundary conditions, (\ref{c6}) changes
the leading behavior (\ref{25}) and not only the next-to-leading behavior.
In fact, (\ref{c6}) permits to change (\ref{25})\ by

\begin{equation}
g_{tt}\simeq -\frac{r^{2}}{l^{2}}\log (r)-\frac{r^{2}}{l^{2}}+\mathcal{O}%
(\log (r)).
\end{equation}%
and something similar for $g_{\phi \phi }$. This type of boundary conditions
was studied in \cite{Henneaux2} for the case of TMG, where it was shown that
these are consistent with asymptotic conformal symmetry at the boundary.
Explicit solutions obeying these asymptotic conditions were analyzed in
Refs. \cite{GAY, Clement}. In the next section we will review this type of
solutions in the case of NMG.

\section{Non-linear solution}

\subsection{Deformation of BTZ\ solution}

Consider first the extremally rotating BTZ\ solution%
\begin{equation}
ds_{e\text{BTZ}}^{2}=-N^{2}(r)dt^{2}+\frac{dr^{2}}{N^{2}(r)}+r^{2}\left(
N_{\phi }(r)dt-d\phi \right) ^{2}  \label{btz}
\end{equation}%
with $r\in \mathbb{R}_{\geq 0}$, $t\in \mathbb{R}$, $\phi \in \lbrack 0,2\pi
)$, where%
\begin{equation}
N^{2}(r)=\frac{r^{2}}{l^{2}}-4M+\frac{4M^{2}l^{2}}{r^{2}},\quad N_{\phi }(r)=%
\frac{2Ml}{r^{2}}.  \label{btz2}
\end{equation}

For $M>0$ this metric exhibits an event horizon at $r=\sqrt{2M}l$. Being an
Einstein space with negative cosmological constant in three dimensions, BTZ\
geometry is locally equivalent to AdS$_{3}$, and it is asymptotically AdS$%
_{3}$ in the sense of \cite{BH}. The parameter $M$ in (\ref{btz})-(\ref{btz2}%
) in the case of General Relativity corresponds to the mass and angular
momentum of the extremally rotating black hole. In the case of TMG at the
chiral point $\mu =1/l$, in contrast, the mass of such a solution is zero,
and the same happens in NMG\ at $m^{2}l^{2}=1/2$.

Now, consider a deformation of (\ref{btz})-(\ref{btz2}) of the form 
\begin{equation}
ds^{2}=ds_{e\text{BTZ}}^{2}+H_{ab}dz^{a}dz^{b}  \label{monster}
\end{equation}%
where $z^{\pm }$ are the coordinates introduced before, with $a,b=+,-$.
Here, $H_{ab}$ are three functions of the coordinates $\phi ,$ $t,$ and $r$.
The large $r$ expansion of $H_{ab}$ determines the asymptotic boundary
conditions.

It turns out that an exact solution to NMG at the point $m^{2}l^{2}=1/2$ ($%
c=0$) is given by the particular deformation%
\begin{equation}
H_{ab}(r)=l^{4}\delta _{a}^{+}\delta _{b}^{+}\left( k_{0}+k_{2}r^{2}\right)
\log \left( \frac{r^{2}-2Ml^{2}}{2Ml^{2}}\right)  \label{monster2}
\end{equation}%
with $k_{0}$ and $k_{2}$ being two arbitrary constant.

Geometry (\ref{btz})-(\ref{monster2}) is not conformally flat, so it is not
an Einstein manifold. Still, it presents constant curvature invariants that
only depend on $l$. This geometry presents a curious geodesic structure at
the radius \thinspace $r=\sqrt{2M}l$, where the extremal BTZ\ geometry
presents its horizon. In general, function (\ref{monster2}) diverges at $r=%
\sqrt{2M}l$, except in the special case $k_{0}=-2Ml^{2}k_{2}$ we will refer
to later.

Solutions similar to (\ref{btz})-(\ref{monster2}) exist for NMG coupled to
TMG at the point $c_{-}=0$. For the theory with $1/\mu =0$, provided it is
parity-even, (\ref{btz})-(\ref{monster2}) remains a solution if one changes $%
N_{\phi }\rightarrow -N_{\phi }$ and $z^{\pm }\rightarrow z^{\mp }$. Metrics
(\ref{btz})-(\ref{monster2}) were also studied in Ref. \cite{GAY, Clement},
and solutions locally equivalent to them appeared already in \cite{Gibbons,
Waves}.

\section{Conserved charges}

\subsection{Auxiliary field formulation}

In this section, we will compute the conserved charges associated to
solution (\ref{btz})-(\ref{monster2}). To do so, we will first write the
quasilocal stress-tensor, which first requires the introduction of suitable
boundary terms in the action. In order to write down the boundary terms for
NMG, it is convenient to rewrite the theory in a different way. In fact,
there exists another action, other than (\ref{NMG1}), from which equations (%
\ref{eom})-(\ref{eom2}) may de derived. This amounts to introduce a
symmetric rank-two auxiliary field $f_{\mu \nu }$ and consider the
alternative action%
\begin{equation}
S_{\text{A}}=\frac{1}{16\pi }\int_{\Sigma }{d^{3}x\sqrt{-g}}\left( R+{f^{\mu
\nu }G_{\mu \nu }-{\frac{1}{4}m^{2}(f_{\mu \nu }f^{\mu \nu }-f^{2})}}\right)
,  \label{alternativa}
\end{equation}%
where, again, $G_{\mu \nu }\ $is the Einstein tensor made out of metric $%
g_{\mu \nu }$, while $f_{\mu \nu }$ is a non-dynamical field. After varying
with respect to field $f_{\mu \nu }$ we find the equation $f_{\mu \nu
}=(2/m^{2})S_{\mu \nu }$, and plugging this back into action (\ref%
{alternativa}) we recover (\ref{NMG1}).

In general, higher derivatives actions like (\ref{NMG1}) require additional
information to be provided about the variation of the fields at the
boundary. In such cases, it is not enough to fix the variations of the
metric at the boundary, but it is also necessary to specify the variation of
the normal derivative of it. In the case of General Relativity this problem
is solved by the addition of the Gibbons-Hawking term to the action. In the
case of NMG, on the contrary, such boundary term is not enough. The authors
of \cite{HohmTonni} gave a prescription for a variational principle in NMG
based on the second order derivative action (\ref{alternativa}). Following
the criterion of \cite{HohmTonni} it is sufficient to fix the variations of $%
g_{\mu \nu }$ and $f_{\mu \nu }$ at the boundary and add to the action a
suitable boundary term that supplements the Gibbons-Hawking contribution.

\subsection{Boundary terms}

Then, as said, boundary terms $S_{\text{B}}$ are added to action (\ref{NMG1}%
) for the variational principle to be defined in such a way that both the
metric $g_{\mu \nu }$ and the auxiliary field $f_{\mu \nu }$ are fixed on
the boundary $\partial \Sigma $. With this prescription, the boundary action 
$S_{\text{B}}$ reads%
\begin{equation}
S_{\text{B}}=-\frac{1}{8\pi }\int_{\partial \Sigma }d^{2}x\sqrt{-\gamma }%
\left( K+\frac{1}{2}\hat{f}^{ij}(K_{ij}-\gamma _{ij}K)\right) .  \label{Sb}
\end{equation}%
Here, we use the convention that Latin indices $i,j=0,1$ refer to the
coordinates on the constant-$r$ surfaces, while the Greek indices are $\mu
,\nu =0,1,2$ and do include the radial direction $r$ as well. Now, $\gamma
_{ij}$ is the two-dimensional metric induced on $\partial \Sigma $ and $%
K_{ij}$ is the extrinsic curvature, with $K=\gamma ^{ij}K_{ij}$. Matrix
components $\hat{f}^{ij}$ are defined as $\hat{f}^{ij}\equiv
f^{ij}+2f^{r(i}N^{j)}+f^{rr}N^{i}N^{j}$, where $N^{i}$ are the shift
functions of the metric $g_{\mu \nu }$ to be written in
Arnowitt-Deser-Misner (ADM) form

\begin{equation}
ds^{2}=N^{2}dr^{2}+\gamma _{ij}(dx^{i}+N^{i}dr)(dx^{j}+N^{j}dr),
\end{equation}%
with the radial lapse function $N^{2}$.

The first term in (\ref{Sb}) is of course the Gibbons-Hawking term. The
other two terms come from the higher-curvature terms of NMG. Notice that in (%
\ref{Sb}) the field $\hat{f}^{ij}$ couples to the Israel tensor $%
K_{ij}-\gamma _{ij}K$ in the same manner as the field $f^{\mu \nu }$ couples
to the Einstein tensor in the bulk action (\ref{alternativa}). Besides,
pushing this analogy forward, one may suggest to supplement (\ref{Sb}) with
pure boundary terms of the form 
\begin{equation}
S_{\text{C}}=\frac{1}{8\pi }\int_{\partial \Sigma }d^{2}x\sqrt{-\gamma }(m
_{0}+m _{1}\ \hat{f}+m _{2}(\ \hat{f}^{2}-\ \hat{f}_{ij}\hat{f}^{ij})+...),
\label{Sc}
\end{equation}%
with adequate constant mass coefficients $m _{i}$, and $\ \hat{f}\equiv \hat{%
f}^{ij}\gamma _{ij}$. In fact, these terms are in general needed to
regularize the action and define finite conserved charges, for instance in
the context of holographic renormalization \cite{HohmTonni, GiribetLeston,
Goya1, Goya2}. Here, because of the special properties of the theory at the
point $c=0$, we will not need to introduce such terms in the action to
regularize the charges. This is because for the theory on AdS$_{3}$ it is
required to introduce a counterterm like%
\begin{equation}
S_{\text{C}}\propto \ c\ \int d^{2}x\sqrt{-\gamma }  \label{upa}
\end{equation}%
which in this case vanishes. See \cite{HohmTonni, GiribetLeston, Goya1,
Goya2} for a discussion on counterterms and the different choices of
asymptotic conditions.

\subsection{Quasilocal stress-tensor and charges}

Now, having introduced the boundary terms, let us analyze the definition of
the Brown-York tensor in NMG as done in Ref. \cite{HohmTonni}. This tensor
is defined by varying the full action $S=S_{A}+S_{B}+S_{C}$ with respect to
the metric $\gamma ^{ij}$, namely%
\begin{equation}
T_{ij}=\frac{2}{\sqrt{-\gamma }}\frac{\delta S}{\delta \gamma ^{ij}}_{|r=%
\text{const}},  \label{Tij}
\end{equation}%
and then taking the limit $r\rightarrow \infty $. The explicit form of $%
T_{ij}$ is cumbersome and can be found in \cite{HohmTonni}\thinspace . It
involves the fields $\gamma _{ij}$, $f_{ij}$, and its derivatives

Conserved charges are then defined in terms of integrals \cite{BrownYork}%
\begin{equation}
Q[\xi ]=\int ds\ u^{i}T_{ij}\xi ^{j},  \label{carga}
\end{equation}%
where $ds$ is the line element of the constant-$t$ surfaces at the boundary, 
$u$ is a unit vector orthogonal to the constant-$t$ surfaces, and $\xi $ is
the Killing vector that generates the isometry on $\partial \Sigma $ to
which the charge is associated. In the case of the mass, the components of
this vector could be $\xi ^{i}=N_{t}u^{i}$, where the lapse function $N^{t}$
is the lapse function of the two-dimensional metric induced at the boundary
expressed in the ADM decomposition.

Let us compute the charges associated to solutions (\ref{btz})-(\ref%
{monster2}). The action of the theory evaluated at these solutions at large $%
r$ diverges like $\sim c\ r^{2}+\mathcal{O}(1)$, which in this case is
finite in virtue of (\ref{aster}). The conserved charge associated to vector 
$\xi =N_{t}u$ can be shown to behave like%
\begin{equation*}
Q[N_{t}u]=\lim_{r\rightarrow \infty }\frac{2(k_{0}+2Ml^{2}k_{2})}{%
1+l^{2}k_{2}\log (r^{2}/(2Ml^{2}))};
\end{equation*}%
which tends to zero if $k_{2}\neq 0$ while it tends to $2k_{0}$ if $k_{2}=0$%
. However, this is not the definition of quasilocal energy we want for $%
k_{2}\neq 0$ configurations. Instead, we prefer to define the mass with
respect to the boundary Killing vector $\xi =\partial _{t}$. The mass
associated to it reads 
\begin{equation}
Q[\partial _{t}]=2(k_{0}+2Ml^{2}k_{2}).  \label{M}
\end{equation}

For $k_{2}=0$ this result coincides with the result $Q[\partial _{t}]=2k_{0}$
found in \cite{GiribetLeston} for a particular case of this geometry, while
for $k_{2}\neq 0$ this gives a new finite contribution to the mass that
happens to be proportional to the parameter $M$ of the extremal BTZ
solution. Notice that if $k_{2}=0$ then $N_{t}u$ tends to $\partial _{t}$
when $r$ tends to infinity.

The angular momentum, on the other hand, being the charge associated to
Killing vector $\partial _{\phi }$, is given by%
\begin{equation}
Q[\partial _{\phi }]=2(k_{0}+2Ml^{2}k_{2})l.  \label{J}
\end{equation}%
Again, for $k_{2}=0$ one reobtains the result of previous computations \cite%
{GiribetLeston}, while it receives a correction when the $\mathcal{O}%
(r^{2}\log (r))$ terms is included. This means that solutions have mass
equal to angular momentum for all values of $k_{0}$, $k_{2}$, and $M$.

It is a remarkable property of NMG\ at the critical point $m^{2}l^{2}=1/2$ (%
\textit{i.e.} $c=0$) that non-linear solutions exhibiting asymptotic
behavior of this kind happen to have finite conserved charges.

It is worth mentioning that the two terms contributing to (\ref{M}) and (\ref%
{J}) come from different terms in the large $r$ expansion in (\ref{monster2}%
); while the piece that depends on $k_{0}$ comes from the $\mathcal{O}(\log
(r))$, it being consistent with Log-gravity boundary conditions, the second
piece comes from the new $\mathcal{O}(r^{2}\log (r))$ dependence. It is
notable that the latter depends both on $k_{2}$ and $M$.

There is a special case for which the deformation (\ref{monster2}) vanishes
at $r^{2}=2Ml^{2}$; namely, when $k_{0}=-2Ml^{2}k_{2}$. With this choice of
parameters the solution exhibits special features; for instance, the
effective potential of geodesics does not diverge. Remarkably, in this case
conserved charges (\ref{M}) and (\ref{J}) vanish.

\begin{equation*}
\end{equation*}

This work has been supported by UBA, CONICET, and ANPCyT. The authors are
grateful to St\'{e}phane Detournay and Guillem P\'erez-Nadal for
discussions. G.G. thanks Pontificia Universidad Cat\'{o}lica de Valpara\'{\i}%
so for the hospitality during his stays, where part of this work was done.


\begin{thebibliography}{99}
\bibitem{NMG} E. Bergshoeff, O. Hohm, P. Townsend, \textit{Phys. Rev. Lett.} 
\textbf{102} (2009) 201301.

\bibitem{NMG2} E. Bergshoeff, O. Hohm, P. Townsend, \textit{Phys. Rev.} 
\textbf{D79} (2009) 124042.

\bibitem{TMG} S. Deser, R. Jackiw, S. Tempelton, \textit{Annals Phys.} 
\textbf{140} (1982) 372.

\bibitem{TMG2} S. Deser, R. Jackiw, S. Tempelton \textit{Phys. Rev. Lett.} 
\textbf{48} (1982) 975.

\bibitem{GrumillerHohm} D. Grumiller and O. Hohm, \textit{Phys. Lett.} 
\textbf{B686} (2010) 2640905.

\bibitem{GrumillerJohansson} D. Grumiller, N. Johansson, \textit{JHEP} 
\textbf{0807} (2008) 134.

\bibitem{BH} J. D. Brown, M. Henneaux, \textit{Commun. Math. Phys.} \textbf{%
104} (1986) 207.

\bibitem{Malda} J.M. Maldacena, \textit{Adv. Theor. Math. Phys.} \textbf{2}
(1998) 231.

\bibitem{ChiralGravity} Li, W. Song, A. Strominger, \textit{JHEP} \textbf{%
0804} (2008) 082.

\bibitem{CDWW} S. Carlip, S. Deser, A. Waldron, D. Wise, \textit{Phys. Lett.}
\textbf{B666} (2008) 272.

\bibitem{CDWW2} S. Carlip, S. Deser, A. Waldron, D. Wise, \textit{Class.
Quant. Grav.} \textbf{26} (2009) 075008.

\bibitem{GKP} G. Giribet, M. Kleban, M. Porrati, \textit{JHEP} \textbf{0810}
(2008) 045.

\bibitem{LogGravity} A. Maloney, W. Song, A. Strominger, \textit{Phys. Rev.} 
\textbf{D81} (2010) 064007.

\bibitem{GrumillerJohansson2} D. Grumiller, N. Johansson, \textit{Int. J.
Mod. Phys.} \textbf{D17} (2009) 2367.

\bibitem{Henneaux1} M. Henneaux, C. Mart\'{\i}nez, R. Troncoso, \textit{%
Phys. Rev.} \textbf{D82} (2010) 064038.

\bibitem{Zanelli} M. Henneaux, C. Mart\'{\i}nez, R. Troncoso, J. Zanelli, 
\textit{Phys. Rev.} \textbf{D65} (2002) 104007.

\bibitem{OTT} J. Oliva, D. Tempo, R. Troncoso, \textit{JHEP} \textbf{0907}
(2009) 011.

\bibitem{Henneaux2} M. Henneaux, C. Mart\'{\i}nez, R. Troncoso, \textit{%
Phys. Rev.} \textbf{D79} (2009) 081502R.

\bibitem{Sun1} Y. Liu, Y-W. Sun, \textit{JHEP} \textbf{0904} (2009) 106.

\bibitem{Sun2} Y. Liu, Y-W. Sun, \textit{JHEP} \textbf{0905} (2009) 039.

\bibitem{Sun3} Y. Liu, Y-W. Sun, \textit{Phys. Rev.} \textbf{D79} (2009)
126001.

\bibitem{NewBC} G. Comp\`{e}re, W. Song, A. Strominger, arXiv:1303.2662.

\bibitem{BrownYork} J. Brown, J. York, \textit{Phys. Rev.} \textbf{D47}
(1993) 1407.

\bibitem{HohmTonni} O. Hohm, E. Tonni, \textit{JHEP} \textbf{1004} (2010)
093.

\bibitem{GiribetLeston} G. Giribet, M. Leston, \textit{JHEP} \textbf{1009}
(2010) 070.

\bibitem{Goya1} G. Giribet, A. Goya, M. Leston, \textit{Phys. Rev.} \textbf{%
D84} (2011) 066003.

\bibitem{Troncoso} F. Correa, C. Mart\'{\i}nez, R. Troncoso, \textit{JHEP} 
\textbf{1202} (2012) 136.

\bibitem{Goya2} G. Giribet, A. Goya, \textit{JHEP} \textbf{1303} (2013)130.

\bibitem{BTZ} M. Ba\~{n}ados, C. Teitelboim, J. Zanelli, \textit{Phys. Rev.
Lett.} \textbf{69} (1992) 1849.

\bibitem{BTZ2} M. Ba\~{n}ados, M. Henneaux, C. Teitelboim, J. Zanelli, 
\textit{Phys. Rev.} \textbf{D48} (1993)\ 1506.

\bibitem{Gibbons} G. Gibbons, C. Pope, E. Sezgin, \textit{Class. Quant. Grav.%
} \textbf{25} (2008) 205005.

\bibitem{GAY} A. Garbarz, G. Giribet, Y. V\'{a}squez, \textit{Phys. Rev.} 
\textbf{D79} (2009) 044036.

\bibitem{Clement} G. Cl\'{e}ment, \textit{Class. Quant. Grav.} \textbf{26}
(2009) 165002.

\bibitem{NMG4D} E. Bergshoeff, J.. Fern\'{a}ndez-Melgarejo, J. Rosseel, P.
Townsend, \textit{JHEP} \textbf{1204} (2012) 070.

\bibitem{Waves} E. Ay\'{o}n-Beato, G. Giribet, M. Hassa\"{\i}ne, \textit{JHEP%
} \textbf{0905} (2009) 029.

\bibitem{FG} C. Fefferman, C. Graham, \textit{Conformal Invariants, }1985 in%
\textit{\ Elie Cartan et les Math\'{e}matiques d'aujourd'hui} (Asterisque)
95.

\bibitem{Cunliff} C. Cunliff, \textit{JHEP} \textbf{1304} (2013) 141.
\end{thebibliography}
\end{document}